\documentclass[%
 aip,
rsi,
 amsmath,amssymb,
 reprint,%
]{revtex4-1}

\usepackage{adjustbox}

\usepackage{dcolumn}
\usepackage{bm}

\usepackage[utf8]{inputenc}
\usepackage[T1]{fontenc}
\usepackage{mathptmx}
\usepackage{etoolbox}
\usepackage[version=4]{mhchem}
\usepackage{amsmath}
\usepackage{mathrsfs}
\usepackage{multirow}
\usepackage{tabularx}
\usepackage{placeins}
\usepackage{float}
\usepackage{subcaption}
\usepackage{soul}
\usepackage{acro}

\usepackage[normalem]{ulem}

\usepackage{tikz}
\usetikzlibrary{shapes.geometric, arrows, arrows.meta, positioning, fit}

\DeclareMathAlphabet{\mathcal}{OMS}{cmsy}{m}{n}

\newcommand{\argmax}[1]{\underset{#1}{\operatorname{\text{arg}}\,\operatorname{\text{max}}}\;}

\usepackage{caption}
\newcommand{\orcidicon}{\includegraphics[width=8pt]{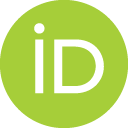}}
\newcommand{\orcidauthor}[1]{%
    \href{https://orcid.org/#1}{\orcidicon}%
}

\usepackage{hyperref}

\hypersetup{
  colorlinks   = true, 
  urlcolor     = blue, 
  linkcolor    = black, 
  citecolor   = red 
}

\makeatletter
\def\@email#1#2{%
 \endgroup
 \patchcmd{\titleblock@produce}
  {\frontmatter@RRAPformat}
  {\frontmatter@RRAPformat{\produce@RRAP{*#1\href{mailto:#2}{#2}}}\frontmatter@RRAPformat}
  {}{}
}%

\makeatother

\DeclareAcronym{BO}{
  short=BO,
  long=Bayesian optimization,
}

\DeclareAcronym{BOIS}{
  short=\texttt{BOIS},
  long=Bayesian Optimization for Ion Steering,
}

\DeclareAcronym{ISAC}{
  short=ISAC,
  long=Isotope Separator and Accelerator,
}

\DeclareAcronym{TRIUMF}{
  short = TRIUMF,
  long  = Canada's National Laboratory for Particle and Nuclear Physics,
}

\DeclareAcronym{ML}{
  short = ML,
  long  = machine learning
}

\DeclareAcronym{GP}{
  short = GP,
  long  = Gaussian process
}
\DeclareAcronym{fc}{
  short = FC,
  long  = Faraday cup
}
\DeclareAcronym{RPM}{
  short = RPM,
  long  = Rotatory Profile Monitor
}
\DeclareAcronym{LPM}{
  short = LPM,
  long  = Linear Profile Monitor
}
\DeclareAcronym{OLIS}{
  short = OLIS,
  long  = Offline Ion Source
}
\DeclareAcronym{RL}{
  short = RL,
  long  = Reinforcement Learning
}
\DeclareAcronym{DRL}{
  short = DRL,
  long  = Deep Reinforcement Learning
}

\DeclareAcronym{HLA}{
  short = HLA,
  long = High Level Applications
}

\DeclareAcronym{NN}{
  short = NN,
  long = Neural Networks
}

\DeclareAcronym{MTBO}{
  short = MT-BO,
  long = Multi-task Bayesian Optimization
}

\DeclareAcronym{AF}{
  short = AF,
  long = acquisition function
}

\DeclareAcronym{RIB}{
  short = RIB,
  long = Rare Isotobe Beams
}

\DeclareAcronym{UCB}{
  short = UCB,
  long = Upper Confidence Bound
}

\DeclareAcronym{EI}{
  short = EI,
  long = Expected Improvement
}

\begin{document}

\newcommand{\BOname}{BOIS} 

\preprint{AIP/123-QED}

\title[Bayesian Optimization for Ion Beam Centroid Correction]{Bayesian Optimization for Ion Beam Centroid Correction}

\author{\textbf{E. Ghelfi} \orcidauthor{0009-0004-3657-2155}}
  \affiliation{TRIUMF, Vancouver, BC, V6T 2A3, Canada}
 \affiliation{
School of Physics and Astronomy, University of Edinburgh, Edinburgh, Lothian, EH9 3FD, UK}

\author{\textbf{A. Katrusiak} \orcidauthor{0009-0005-2088-2033}}%
\thanks{ E.G. and A.K. have contributed to this paper equally and are listed alphabetically}
\affiliation{%
TRIUMF, Vancouver, BC, V6T 2A3, Canada}
\affiliation{Department of Physics, Engineering Physics \& Astronomy,
Queen's University, Kingston, ON, K7L 3N6
, Canada}

\author{R. Baartman \orcidauthor{0000-0002-4007-1220}}
 \homepage{ http://lin12.triumf.ca}
\affiliation{%
TRIUMF, Vancouver, BC, V6T 2A3, Canada}
\affiliation{Department of Physics and Astronomy, University of Victoria, Victoria, BC, V8W 2Y2, Canada}

\author{W. Fedorko \orcidauthor{0000-0002-5138-3473}}
\affiliation{%
TRIUMF, Vancouver, BC, V6T 2A3, Canada}

\author{O. Kester \orcidauthor{0000-0002-1809-5031}}
\affiliation{%
TRIUMF, Vancouver, BC, V6T 2A3, Canada}
\affiliation{Department of Physics and Astronomy, University of Victoria, Victoria, BC, V8W 2Y2, Canada}

\author{G. Kogler Anele \orcidauthor{0009-0005-4614-8160}}%
\affiliation{%
TRIUMF, Vancouver, BC, V6T 2A3, Canada}
 \affiliation{Department of Physics \& Astronomy, University of British Columbia, Vancouver, BC, V6T 1Z1, Canada}

\author{O. Shelbaya \orcidauthor{0000-0003-1796-3965}}
\email{oshelb@triumf.ca}
\affiliation{%
TRIUMF, Vancouver, BC, V6T 2A3, Canada}
             
\author{D. Tanyer \orcidauthor{0009-0005-7009-0106}}%
\affiliation{%
TRIUMF, Vancouver, BC, V6T 2A3, Canada}
\affiliation{Department of Physics and Astronomy, University of Waterloo, Waterloo, ON, N2L 3G1, Canada}

\date{\today}

\begin{abstract}
An activity of the TRIUMF automatic beam tuning program, the Bayesian optimization for Ion Steering (\texttt{BOIS}) method has been developed to perform corrective centroid steering of beams at the TRIUMF ISAC facility. \texttt{BOIS} exclusively controls the steerers for centroid correction after the transverse optics have been set according to theory. The method is fully online, easy to deploy, and has been tested in low energy and post-accelerated beams at ISAC, achieving results comparable to human operators. \texttt{scaleBOIS} and \texttt{boundBOIS} are na\"{i}ve proof-of-concept solutions to preferably select beam paths with minimal steering. Repeatable and robust automated steering reduces reliance on operator expertise and operational overhead, ensuring reliable beam delivery to the experiments, and thereby supporting TRIUMF's scientific mission.

\end{abstract}

\maketitle

\section{\label{sec:intro}Introduction}

\begin{figure*}[htb]
    \centering
     \includegraphics[width=0.7\textwidth]{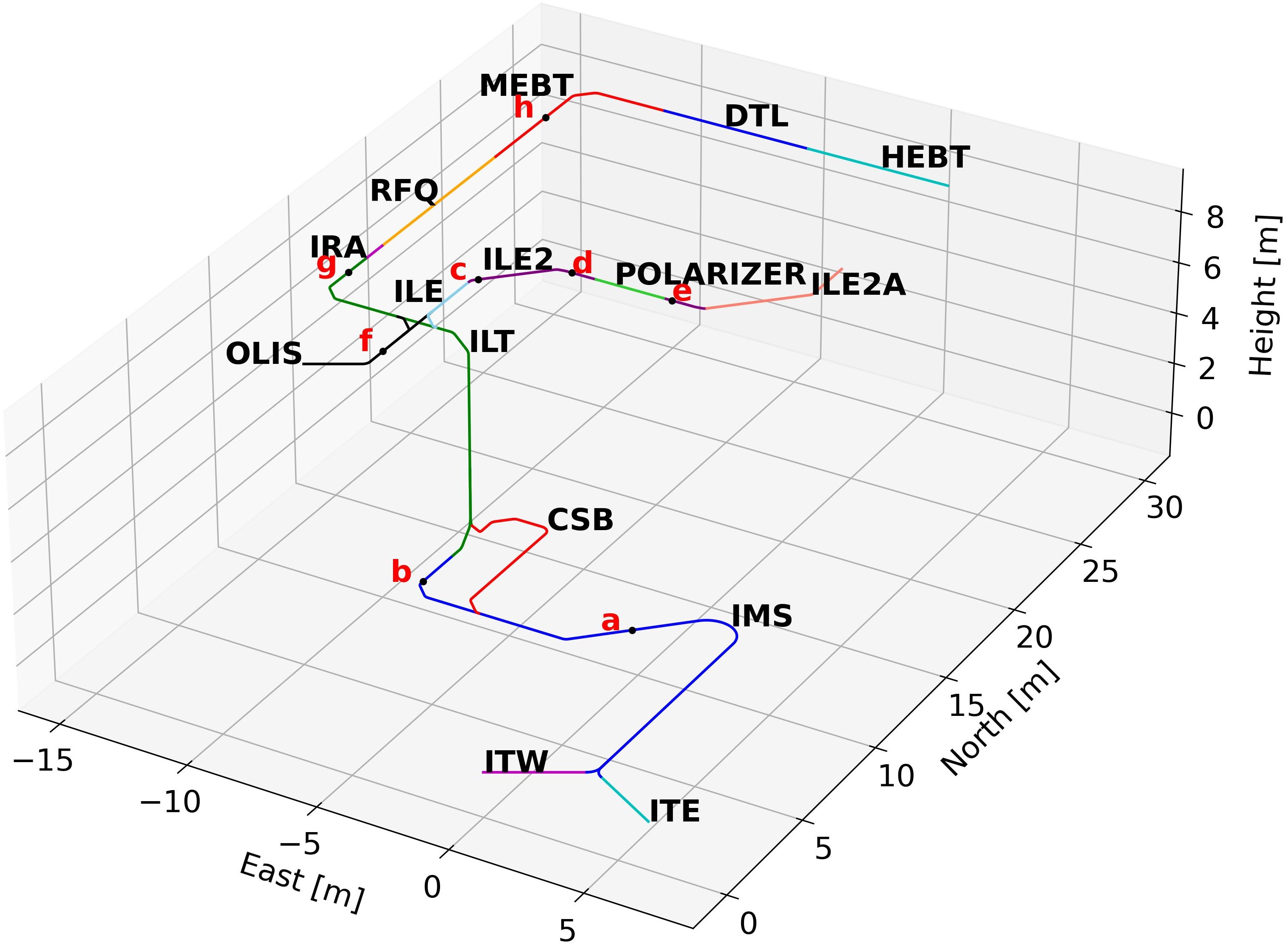}
    \caption{\label{fig:ISAC}Some of the ISAC beamlines. Breakpoints used for testing are shown in red, which will be used in \autoref{sec:Results}.}
\end{figure*}

Rare Isotope Beam (RIB) facilities enable the study of diverse nuclear properties and interaction cross sections, serving studies in material science, fundamental symmetries, nuclear structure, and nuclear astrophysics \cite{dilling2014ariel}. TRIUMF has embarked on a program to implement high level applications (HLA) and automatic beam tuning to raise the efficiency of RIB delivery for maximum benefit of the experiments served by TRIUMF beams. 

In this paper we present the results of a \ac{BO} algorithm for online centroid correction of a non-space-charge dominated RIB, developed and used at TRIUMF's Isotope Separator and Accelerator (ISAC, Fig.~\ref{fig:ISAC}) facility\cite{dilling2014isac, dombsky2014isac}. ISAC produces RIB by proton bombardment of targets \cite{bricault2001production}, which generates various radionuclides through processes that include spallation \cite{russell1990spallation}, fragmentation \cite{lynch1987nuclear}, and to a lesser extent fission \cite{bricault2014rare}. 

At ISAC, beam centroid correction procedures have to date been performed manually: an operator tunes the many steerers throughout the beamline, while monitoring transmission using \acp{fc}, aiming to center the beam and optimize transmission. This large parameter space operation is resource intensive, monopolizing the operator's attention. Thus, within the automatic beam tuning program there  is an interest in exploring algorithmic means, including machine learning methods, to offload this task from the operators, instead providing them with a high-level application capable of performing this optimization. Previously, a deep reinforcement learning (DRL) model was developed by Wang et al. (2021) \cite{Wang2021drl} on a simulation of the beamline, and while successful at tuning in a simulated environment direct translation to on-line tuning was not possible due to the differences between the simulation and real system configuration and operating parameters. Moreover, a DRL model that is completely trained on-line was deemed not viable due to the prohibitively large number of iterations required.  

Providing more tools to operators is an important step in preparing for the ARIEL \cite{dilling2014ariel} era at TRIUMF, which will see the activation of a new network of beam transport and acceleration beamlines, including the new electron linear accelerator (linac), that will drive photofission based, neutron-rich, isotope production. Additionally, a new proton beamline \cite{Rao:2018edq} from TRIUMF’s main cyclotron \cite{bylinskii2014triumf} is being built together with a new proton target station \cite{augusto2022design}. Finally, ARIEL will include the new CANREB facility \cite{ames2018canreb}, with a new high-resolution separator with $M/(\delta M) > 10000$, complementing ISAC RIB mass separation for the two existing ISAC east and west targets \cite{bricault2002triumf} (Figure 1, ITE and ITW). In all, an anticipated tripling of delivered RIB-hours to experimental users is planned. With increasing beam delivery to users comes a necessary increase in operator workload, hence their time and attention will become an ever more crucial resource.

Recent work \cite{Shelbaya2023mcat} has developed parallel-modeling capabilities for machine tuning optimizations. Using the \texttt{TRANSOPTR}\cite{Heighway:1984zf} first order envelope code, the beam transport optics (quadrupoles, dipoles, spherical benders, etc.) and post-accelerators (RFQ\cite{shelbaya2019fast} and DTL\cite{shelbaya2021autofocusing}) can be set using optimization of a digital twin of the system; this is achieved using a high-level application: Model Coupled Accelerator Tuning (MCAT)\cite{Shelbaya2023mcat}. Nevertheless, parallel modeling \textit{only} computes the necessary set-points for optics devices, and does not perform steering optimizations as this would require the knowledge of all potential misalignment and field error sources. The procedure presented here, in tandem with MCAT, increases repeatability and reduces the complexity of the beam tuning procedure. 

\ac{BO} is an excellent candidate to optimize a noisy black-box function which is expensive to evaluate \cite{roussel2024bayesian}. It uses a \ac{GP} as a surrogate model for the objective function and an \ac{AF} to select trial points. After testing the system with the trial points it uses Bayes' theorem to update the GP at every iteration. 

This study demonstrates the ability to perform corrective beam steering across the low-energy electrostatic beamlines, typically working with ion beam currents in the nA range, using  beams from the ISAC Mass Separator\cite{doornbos1997mass} (IMS) radioactive beam targets, in addition to the stable pilot beams from the Offline Ion Source\cite{jayamanna2008off,jayamanna2010multicharge} (OLIS). 

We present the \ac{BOIS} method, which works exclusively with online data for tuning and without prior training. Importantly, it is defined in a context where the optics are set and the remaining optimization problem is steering.

A major finding is that it can match or exceed the capabilities of expert operators in terms of transmission and tuning time. This marks a significant step forward in the automation of beamline and accelerator tuning.

We propose additional model versions to minimize steering to avoid solutions with large transverse excursions. This aligns with the direction of ongoing developments within the automatic beam tuning program. \texttt{boundBOIS} imposes a stricter bound to limit steering to the order of the beam divergence, and \texttt{scaleBOIS} handles a modified current value, altered by a scalar assessing how close the system is to neutral steering. 

As a comparison, at the Paul Scherrer Institut (PSI), bounds have been implemented in terms of safety constraints by Kirschner et al. (2022) \cite{Kirschner2022safeBO} who also propose sub-dividing the AF problem to handle more parameters \cite{Kirschner2019Bayesian}. At the Facility for Rare Isotope Beams (FRIB), Hwang et al. (2022) \cite{Hwang2022} train a neural network on historical data to initially determine the prior GP. Online multi-objective BO has been developed to include the optimization of beam spread in a laser-plasma accelerator \cite{jalas2023multi} or beam parameters exiting a linac \cite{Roussel2021Multiobjective}. Other machine learning algorithms have been applied to accelerator tuning problems granting fully online optimizations, such as combining an extremum seeking algorithm with deep neural networks to tune the AWAKE beamline \cite{scheinker2020}. 

In this paper, the problem of beam centroid correction is firstly described in \autoref{sec:centroids}, followed by an overview of Bayesian optimisation in \autoref{sec:BO}. This feeds into the discussion of the \texttt{BOIS} method in \autoref{sec:bois}, which includes the \texttt{scaleBOIS} and \texttt{boundBOIS} for reducing steering. Results from online tuning are presented for different segments of the low-energy beamlines at ISAC in \autoref{sec:Results}, before concluding this work in \autoref{sec:conclusion}.

\section{\label{sec:centroids}The Beam Centroid Correction problem}

\begin{figure}[ht!]
    \includegraphics[width=0.5\textwidth]{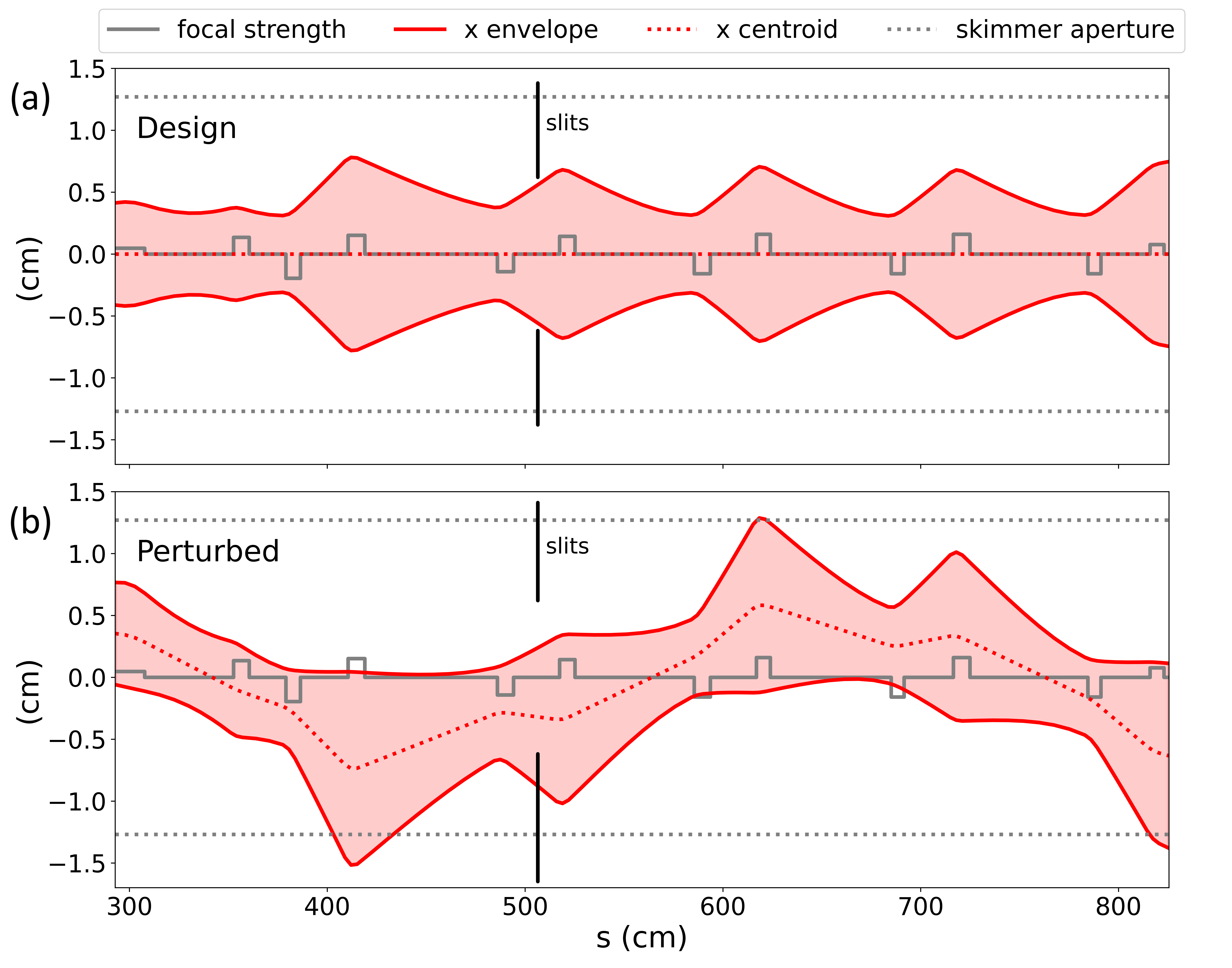}
    
    \caption{The 2~rms x envelope of an example beam transport. (a) shows the designed behavior of the beam, (b) shows the same optics settings with a perturbed beam entering the lattice (shifted: $x=0.35 \text{~cm}, \text{ } x'=0.002\text{~rad}$). Notably, beam loss occurs at slit locations and the skimmer plates. At ISAC, skimmers are apertures (usually 25~mm) that are used for the electrostatic elements to confine the electromagnetic field.}
    \label{fig:transoptr_design}
\end{figure}

The tuning process involves adjusting the various electromagnetic lenses to achieve optimal transport. Beams differ in energy and charge (ionization) state depending on scheduled experiment requirements, target geometry and source type (surface ionization source (SIS), forced electron beam induced arc discharge (FEBIAD), for instance), requiring constant re-tuning. The aim of the research presented here is to deliver a beam with consistent and stable transmission.

A simplified beam transport problem consists of apertures, drift spaces, and lenses, and is not designed to need steering. However, if any misalignments or field errors exist, the effects propagate downstream via the lensing elements, leading to beam loss at the apertures or beam pipes.  To elaborate, quadrupoles have zero-field on axis and they only act to focus the beam, however, the electromagnetic field off-axis is nonzero, which causes a transverse kick of the beam if its centroid is uncentred with respect to the quadrupole axis: this is shown in \autoref{fig:transoptr_design}. 

Steerers correct for this along the beamline, specifically placed at a $\pi/2$ phase difference to cancel the betatron oscillation in position $x$ and momentum $x'$. These change the small angle displacement of the beam, typically on the order of milliradians. 

There are no theoretical steerer settings as they result from unknowns such as element misalignments either fixed or varying due to floor settling  \cite{TRI-BN-19-18}, and variable stray ambient magnetic fields, for instance at the source. These cannot be predicted, or characterized given the current available diagnostics.  Current measurements will furthermore be subject to random noise which affects an iterative algorithm's capability to maintain stability.

The \ac{BOIS}  approach decouples the control of the central trajectory of the beam and its size. Focusing elements (quadrupoles and benders) control both, while steerers control only beam trajectory \cite{larson1977steering}. Once the optics have been set to computed values for transport, the \ac{BOIS} algorithm is used to adjust steerers, tackle centroid correction, and ensure the beam is well centered. This maximizes transmission by avoiding aperture losses and allows the lenses to focus with minimized steering effect.

\section{\label{sec:BO} Bayesian Optimization: a background}

\acl{BO}\cite{Kushner1964ANM,Mockus1978BayesianMethods,Jones1998Optimization} is a probabilistic-model based optimization strategy for black-box objective functions that are expensive to evaluate. In this case the objective function is the current at a \ac{fc} downstream of a set of steerers, given their values. 

The goal of \ac{BO} is to identify the argument \(\mathbf{x}^*\) that produces a global optimum of the objective function \(f\):
\begin{equation} \label{eq:argmax}
    \mathbf{x}^*=\argmax{\mathbf{x}\in \mathcal{X}}f(\mathbf{x}),
\end{equation}

where in general
\(f\) is a black-box function in the sense that its functional form, derivatives, convexity, and linearity properties are all inaccessible a priori.

The only way one can extract information about the objective function is by evaluating it at a given point. These evaluations are expensive and naturally corrupted by noise.  BO is rooted in Bayesian inference and builds a mathematical surrogate that approximates the objective function while being faster and less expensive to evaluate. 

 A \ac{GP} model is often chosen as the surrogate. A \ac{GP} can be thought of as a distribution over all functions. When the surrogate function is evaluated at a finite set of inputs, its values are assumed to form a multi-variate Gaussian distribution. This implies that a mean value and its associated uncertainty can be computed for any input for the surrogate model. The prior distribution over the \ac{GP} is initially uninformative before any data is used to update it. As new data is added, Bayes' rule is used to compute the posterior distribution, which then becomes the prior for the next iteration \cite[pp. 108--111]{rasmussen2006GPforML}.

At each iteration, BO is concerned with intelligently selecting the next point at which to evaluate the real objective function to update the surrogate distribution. This is done by fitting an \ac{AF} on the surrogate to select the next sampling point to balance the exploitation of the expected maxima and exploration in areas of higher uncertainty. 

\begin{figure}[h!]
    \centering
    \includegraphics[width=0.5\textwidth]{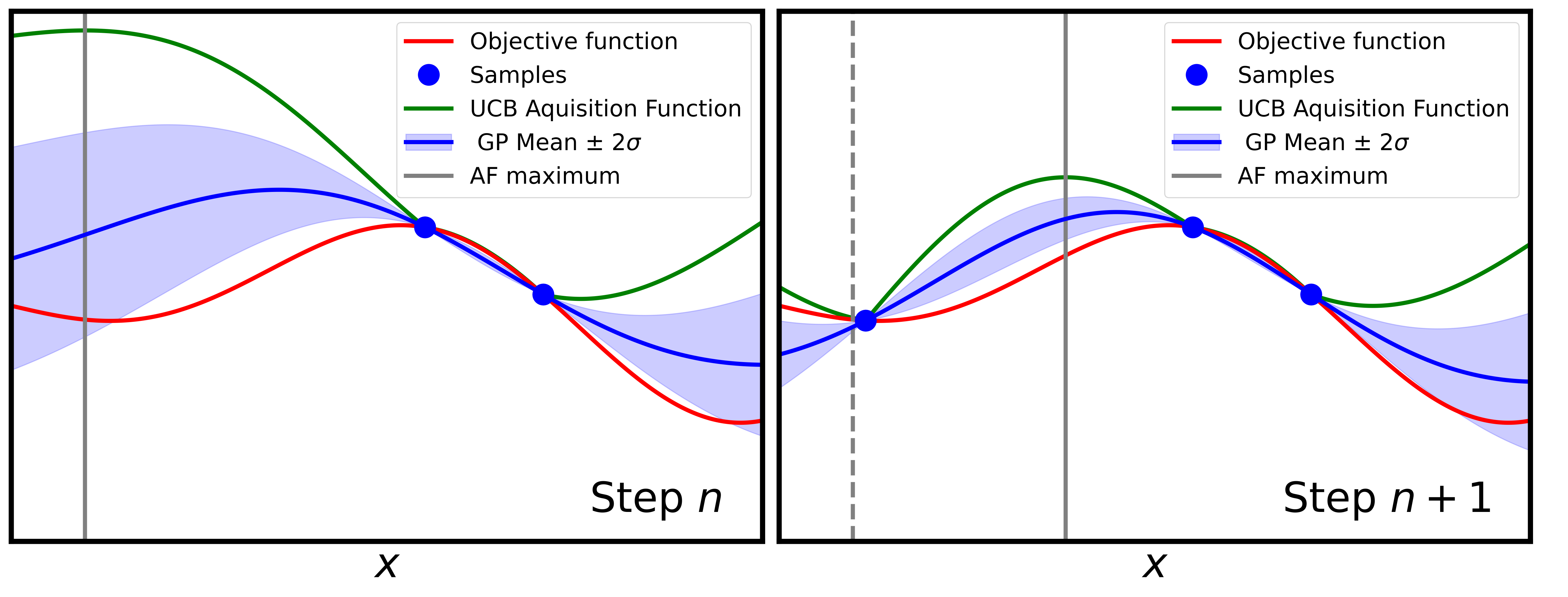}
    \caption{\ac{BO} seeks to use an \ac{AF} (green) to sample (blue dots) an unknown objective function of interest (noisy red) and create a probabilistic model (blue line and shaded area for 2$\sigma$ confidence bounds). The cartoon shows two consecutive steps: the maximum of the \ac{AF} at step $n$ guides the next sampling point at step $n+1$.}
    \label{fig:bo}
\end{figure}

\autoref{fig:bo} depicts a cartoon 1D problem showcasing two consecutive iterations of a BO algorithm, with the acquisition of new data and posterior distribution updates.

The acquisition of new data updates the posterior distribution over the GP function space in the BO algorithm.

\subsection{\label{sec:Gaussian}Gaussian Processes}

A \ac{GP} is composed of random variables, and the joint distribution of any finite subset of these variables is Gaussian, written as\cite[p. 13]{rasmussen2006GPforML}:
\begin{equation}
    \mathbf{f(x)}= [f(x_{1}),f(x_{2}),...,f(x_{N})]^{T} \sim \mathcal{N}(\mu(\mathbf{x}), \mathbf{k(x,x')}),
\end{equation}
where $\mu(\mathbf{x})$ is the mean function and $\mathbf{k(x,x')}$ is the covariance (kernel) function of the normal distribution $\mathcal{N}$. 

The choice of kernel function $\mathbf{k(x,x')}$ is crucial in defining the properties of the GP; it encodes the assumptions about the objective we are learning. Assumptions such as smoothness and length-scale of the correlations between points. A typical choice of kernel function is the Matérn kernel\cite{matern1960kernel}. 

The selection of a prior distribution for the GP hyper-parameters, such as the length scale in the covariance function, reflects our prior beliefs about the underlying function's properties. For instance, by placing a prior distribution on the length scale parameter we can express prior beliefs about the smoothness of the objective function, and avoid over-fitting. These are updated at each iteration. 
The probability distributions of the objective function output is updated using Bayes' rule each time new data is added \cite[pp. 18--19,108--111]{rasmussen2006GPforML}. 

The object of Gaussian process regression is to model the objective function $f(\textbf{x})$, as a GP by updating the data used and performing a Bayesian regression. This is to say that we model the objective function as
\begin{equation}
  p(\mathbf{f(x)}|\mathcal{D})=\mathcal{G}\mathcal{P}\left(\mathbf{f};\mathbf{\mu(x)}_{\mathbf{f}|\mathcal{D}}, \mathbf{k(x, x')}_{\mathbf{f}|\mathcal{D}}\right),  
\end{equation}
where $\mathcal{D}$ is our data set,
\begin{equation}
    D = \{(\mathbf{x}_i,y_i)\}^{N}_{i=1},
\end{equation}
with $N$ points. 

\subsection{Acquisition functions}
Instead of trying to directly optimize the unknown function ${f(\mathbf{x})}$ by finding the maximum of the GP, BO finds a maximum in the \ac{AF} to decide the next point of evaluation. An \ac{AF} balances searching unexplored regions and exploiting known maximal regions in $\mathbf{f(\mathbf{x})}$. 

\ac{AF}s can be either analytic or based on Monte Carlo sampling, which can be more efficient with large dimensional problems and can provide multiple trial points for batched optimization. Here we use an analytic \ac{AF} for simplicity.

Additionally, this provides us with an explicit expression for maximization based on the statistics of the GP posterior. Popular analytic \ac{AF}s are \ac{EI} and \ac{UCB}. 

The \ac{UCB} \ac{AF} \cite{srivanas2010ucb} is defined as: 
\begin{equation} \label{eq:UCB}
    UCB(\mathbf{x}) = \mu(\mathbf{x}) + \sqrt{\beta}\sigma(\mathbf{x}).
\end{equation}

\ac{UCB} explicitly balances exploitation and exploration with a parameter $\beta$: where $\beta<1$ favors exploration of the known maxima and $\beta>1$ favors exploration regions with higher uncertainty.

\begin{figure*}[hbt]
    \centering
    \includegraphics[width=1\textwidth]{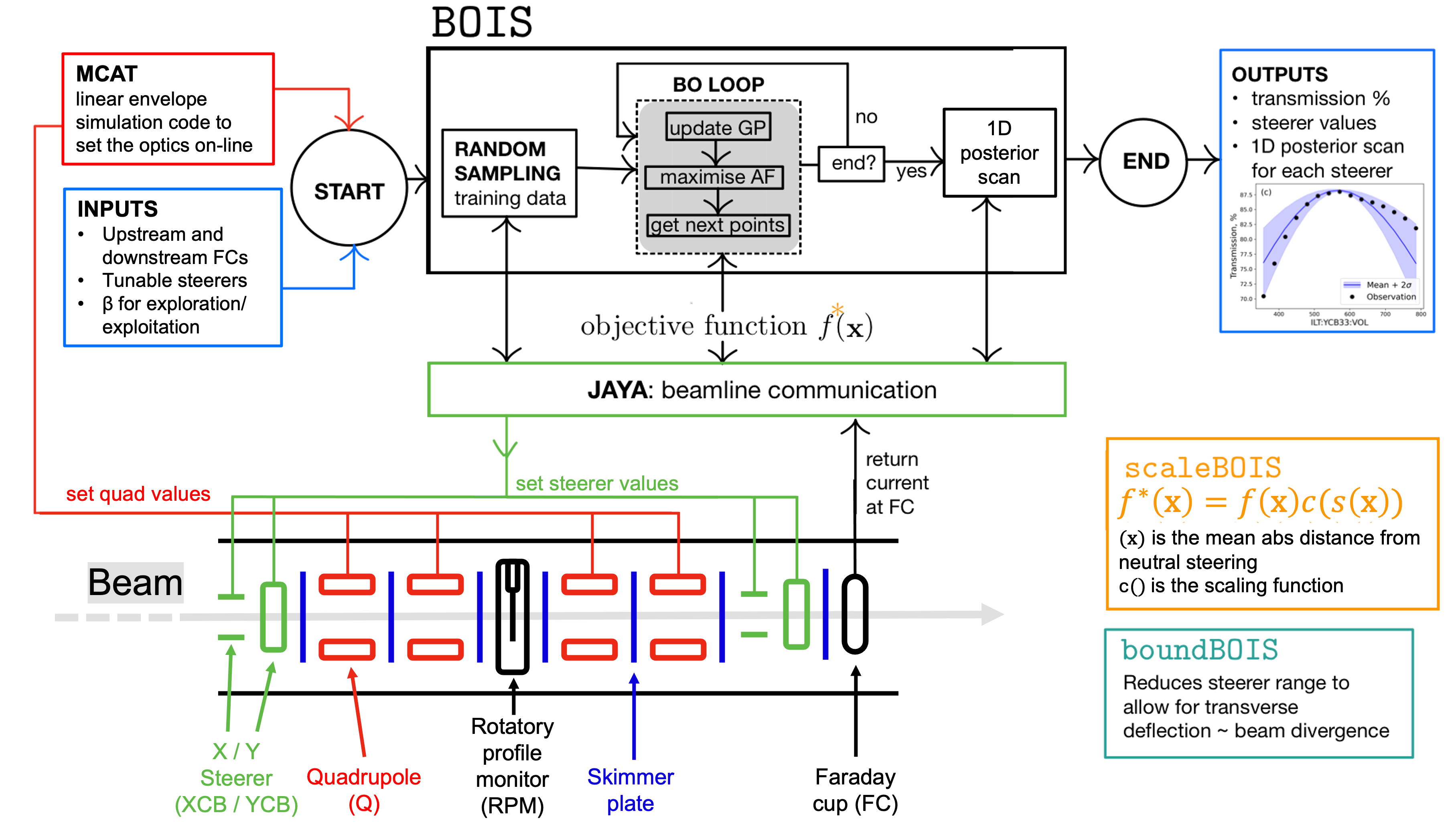}
    \caption{Framework showcasing the intended use of \BOname.}
    \label{fig:BOIS_sketch}
\end{figure*}

\section{\label{sec:bois} \texttt{BOIS}: Bayesian Optimization for Ion Steering}

\texttt{BOIS}, Bayesian Optimization for Ion Steering, is a tool to deploy BO for corrective centroid optimization of radioactive beams, developed and tested at TRIUMF's ISAC facility. Using the methods outlined in \autoref{sec:BO} we solve the problem of beam centroid correction, as laid out in \autoref{sec:centroids}, with steering voltages as parameters and \ac{fc} current as the objective function. There is no distinction made between horizontal and vertical steering. 

The method presented herein utilizes the \ac{BO} and \ac{GP} landscape within the \texttt{BOTORCH} \cite{balandat2020botorchframeworkefficientmontecarlo} and \texttt{Gpytorch} \cite{Gardner2018Gpytorch:Acceleration} libraries in \texttt{python}, which are open-source.

\ac{UCB}  (eq. \ref{eq:UCB}) can selectively, and somewhat aggressively pursue exploration away from the known maxima by utilizing a higher $\beta$ value. In comparison, the exploration-exploitation parameter in \ac{EI} can emphasize exploration only in the context of a high prior mean. The nature of the tuning problem results in multiple possible solutions in a large parameter space, hence fixating on an anticipated maximum is impractical. Moreover, \ac{EI}'s performance is limited if large regions of input space have zero probability of improvement \cite{ament2024unexpected}, which we have noticed as well and is the case here. \ac{EI} has been unreliable in our tests due to the majority of the parameter space giving no transmission.  We found that a slight focus on exploration is suitable for our problem, with $\beta \in [2, 5]$. 

We have chosen the Matérn kernel \cite{matern1960kernel} with a smoothness parameter $\nu =\frac{5}{2}$ because of the flexible application to many physical processes, and found that it achieved higher transmission in fewer iterations compared to Matérn kernels with other smoothness parameters as well as a Gaussian kernel. Rasmussen and Williams (2006) \cite{rasmussen2006GPforML} explain (pp. 83--84) that while the squared exponential kernel is the most widely used kernel within some fields due to its infinite differentiability, Stein  recommends \cite{stein2012interpolation} the Matérn class since the squared exponential is unrealistic for physical processes. The Matérn kernel with $\nu = \frac{5}{2}$ held robustly in all beamlines tested. 

A GP with a $\nu$ Matérn kernel is in general $ \lceil \nu \rceil -1 $ times differentiable. It is standard measure to fix the $\nu$ parameter and optimize the lengthscale parameter due to the computational infeasibility and drastic function changes with changes of $\nu$ \cite[pp. 84--85]{rasmussen2006GPforML}. 
A simple interpretation of the length scale is the distance between two points in the input space which yield significantly different function values
\cite[pp. 14--15]{rasmussen2006GPforML}. 

The hyper parameters of the GP, including the lengthscale parameter of the kernel are optimized to, at each iteration, maximize the likelihood of the data, which is the probability of observing that data given the model parameters. The lengthscale is given a prior probability distribution of a gamma distribution. 

This distribution restricts inputs to real positive numbers because the lengthscale must be strictly positive. The parameters we have chosen for the distribution embed our belief that the objective function is not very smooth, thus highly sensitive to input changes. We wish to permit some flexibility for smoothness, but emphasize shorter correlations.

\subsection{Configuration}
Given theoretical optics settings, \texttt{BOIS} is a fully self-contained method that requires no previous training information, giving it low startup overhead. Configuration of a run involves selecting the tunable steerers and objective \ac{fc}, optimizer parameters, and initial number of sampling points. The run stops when the optimizer converges to a solution, or at any time by an operator.

Upon first implementation of this tool for optimizing the beam in a given beam line, the elements have to be chosen that are used for the optimization. It is known \cite{Frazier2018BayesOpt} that BO becomes too computationally demanding when working with dimensions $d\ge20$.  With assumptions about the objective one can exploit lower dimensionalities in the problem \cite{Moriconi2020highdimensions,eriksson212021highdimensions}, and this has been shown to work on accelerator problems (\textsc{LineBO} \cite{Kirschner2019highdimensions_safe_fel} and \textsc{SafeLineBO}\cite{Kirschner2022safeBO}). However, this approach is not explored here, and therefore we must segment our problem into bite sized pieces. Ultimately, the location of FCs determines the available breakpoints of the subsections, and FCs are destructive so only one section can be tuned at a time. The presented results use between 4 and 17 steerers at the same time.

The FCs used here sample at 10~Hz and a "measurement" is taken as the average of 20 readings from the FC. The noise present in these readings is on the order of pico amps, and they are calibrated periodically using current sources. 

Literature \cite{morar2021bayesian} suggests that an appropriate number of initial sampling points is $2d$ where $d$ is the number of tunable parameters, i.e. steerers. A lower number is not recommended: while a \ac{GP} and \ac{AF} would be more efficient than random sampling they are also more computationally taxing, and a non-linear multi-dimensional Gaussian Process regression based on a very limited number of samples may not be useful. 

The code was developed using a simulated twin of the ISAC beamlines, using TRANSOPTR \cite{Heighway1981Transoptr}: a beam envelope code which uses the quadratic Hamiltonian of a reference particle in the Frenet-Serret coordinate system and a Runge-Kutta method to numerically evaluate the equations of motion and beam matrix given the beam’s initial conditions and the optical elements it crosses \cite{Baartman2017faa}. Since 1984, the TRIUMF Beam Physics group has extended and adapted TRANSOPTR \cite{baartman2016transoptrchanges} to describe TRIUMF facilities and beamlines. For testing, the optimizer used TRANSOPTR for objective function evaluations (to simulate the real beamline); here, random kicks are added to generate misalignments which the optimized steerers can correct for. 

\subsection{Procedure}
The cartoon in \autoref{fig:BOIS_sketch} shows the procedure to use. After using MCAT to set the quadrupole values, a user needs a configuration file with the names of the steerers to tune, and the \ac{fc} to use to check current. A choice of $\beta$ and number of initial sampling points must also be made. In an optimization step, the optimizer builds/updates a GP model and maximizes the AF (UCB) to get the next sampling points. Steerer values are set and currents at the FC are read back via communication with the EPICS \cite{Dalesio:1992fso} system for beamline control. At ISAC, jaya \cite{jayagitlab} is used to monitor read/write requests to the main control system. The optimization continues until the current increase is not significant. 

At this point, the performance of the optimizer can optionally be assessed via 1D posterior scans: each of the steerers (dimensions) is scanned around a neighborhood of the optimized value while keeping the rest constant at their optimum values, and this is compared to the GP posterior. At the end of the optimization, the user has a list of optimal steerer values, a current value at the downstream \ac{fc} (and corresponding transmission from the upstream \ac{fc}), as well as the 1D posterior scan plots. These can be then used to assess optimizer performance.

\subsection{\label{sec:constraining} Constraining steering: \texttt{scaleBOIS} and \texttt{boundBOIS} }
TRIUMF-ISAC beamlines are designed \cite{baartman1997ISAC,baartman2003emittance} with transverse geometric acceptances of $200~ \mu {m}$, compared to typical geometric emittances of around $10~  \mu {m}$ and an upper limit nominal value of $50~  \mu {m}$. The large acceptance to emittance ratio means that beam can be badly centered in a given section and still have near 100\% transmission. This is undesirable for a number of reasons: the same tune will perform more poorly if the emittance increases, or if there are small drifts in other parameters; in other words, the tune will not be as robust. As well, badly-aligned beams cannot be easily re-tuned if, for example, it is required to change the spot size on target or through an aperture, since the focusing elements will all steer and thus change the beam centering and quickly decrease the transmission.
Within the problem of steering the beam from one FC to the next, one is posed with multiple objectives: maximise transmission and reduce use of steering. While these objectives are coupled, it is important to prioritise each of them, and a high success in one of the two may result in failing the other - describing a multi-objective problem. 

As a na\"{i}ve, proof-of-concept solution we have modified \texttt{BOIS} to scale down perceived objective function output based on a score calculated from the use of steerers, thus penalizing extreme steerer usage (\texttt{scaleBOIS}). Incorporating this into a model is feasible, while it would be challenging for human operators to manually account for. Additionally, we have implemented more constrained bounds on the input space, which scale with beam energy, to ensure that the steering angle imparted does not exceed the angle on the order of the beam divergence (\texttt{boundBOIS}). 

\subsubsection{\label{sec:scaling}\textnormal{\texttt{scaleBOIS}}: constraint-based perceived beam loss}
Scalarization is a popular approach \cite{roussel2024bayesian, Roussel2021Multiobjective} to multi-objective optimization problems. This is where multiple objectives are reduced to a single scalar value, simplifying the problem to a single-objective optimization. How this reduction takes place is important, and defines the assumed trade-offs between the different objectives. The ultimate form of this approach is to find the Pareto front, which is the set of solutions that are non-dominant in one objective, and therefore find the best compromise \cite{MOHANTY2017pareto}. Our approach simply finds \textit{a} compromise, which is not necessarily Pareto optimal, meaning our \ac{AF} does not consider maximizing the hypervolume in objective-space. At this point, the aim is to simply reduce steering and avoid excessive beam excursions, not find the optimal balance since the ultimate goal is still transmission. Technically all solutions that reach the maximum physical transmission exist on the Pareto front.

Our method involves a nonlinear scalarization of the objective function and scaling factor. This approach optimizes a super-objective function \cite[p. 1--2]{Hwang1979multiobjectivedecisionmaking}, and our new optimization problem looks like:

\begin{equation}
    \mathbf{x}^*=\argmax{\mathbf{x}\in \mathcal{X}}f(\mathbf{x})c(\mathbf{x}),
\end{equation}

where $f$ is the principal objective function (i.e. transmission) and $c$ is the scaling function which we call the centeredness value, $ c \in [0,1]$, where 1 represents a beam with no steering. We calculate the mean, $s$, of the steering deviations from neutral in each steerer, normalized to $\in [0,1]$, which serves as the argument of a weighting function returning a value $c$. We have found best results using a quadratic function:
\begin{equation}
    c(\mathbf{x}) = -p s(\mathbf{x})^2 +1 ,
\end{equation}
for the penalization, where $p \in (0,1] $ is a parameter for the strictness of penalization. In our case we used $p=1/4$ for development, but this number will ultimately depend on the beamline and beam. The choice of a parabolic response to steering is somewhat arbitrary and only reflects our desire to have a simple concave down function to convert the steering deviation to a scaling factor. The nonlinear response is important because the trade-off between these objectives is not constant. Generally, we were most concerned about being lenient to small deviations around zero and increasing intensity gradually. We also explored a Gaussian and circular function, but found them too lenient to large steering.

\subsubsection{\label{sec:bounds_method}\textnormal{\texttt{boundBOIS}}: bounded inputs}

Reducing the viable input space was used for \texttt{boundBOIS}. Here we use knowledge about where a region of the input space exists which contains the Pareto front. This excludes solutions with excessive steering. The addition of constraints to favour the objective of low steering is comparable to an \textit{a priori}\cite[p. 250]{Hwang1979multiobjectivedecisionmaking} version of $\varepsilon$-constraint \cite{Haimes1971epsilonconstraint}. 

Steering voltages were limited to a region that induced transverse deflection in the beam on the order of its divergence. This being taken as a $\pm$2~mrad bound on deflection. Effectively this is realized in the model as a limit on steering voltage which scales with beam energy. Limiting the voltage range also reduces the input space to a more useful region, assisting in optimization performance.

\section{\label{sec:Results}Results}

\begin{figure*}[hbt]
    \centering
    \begin{subfigure}{0.34\textwidth}
        \includegraphics[height = 3.2cm, keepaspectratio]{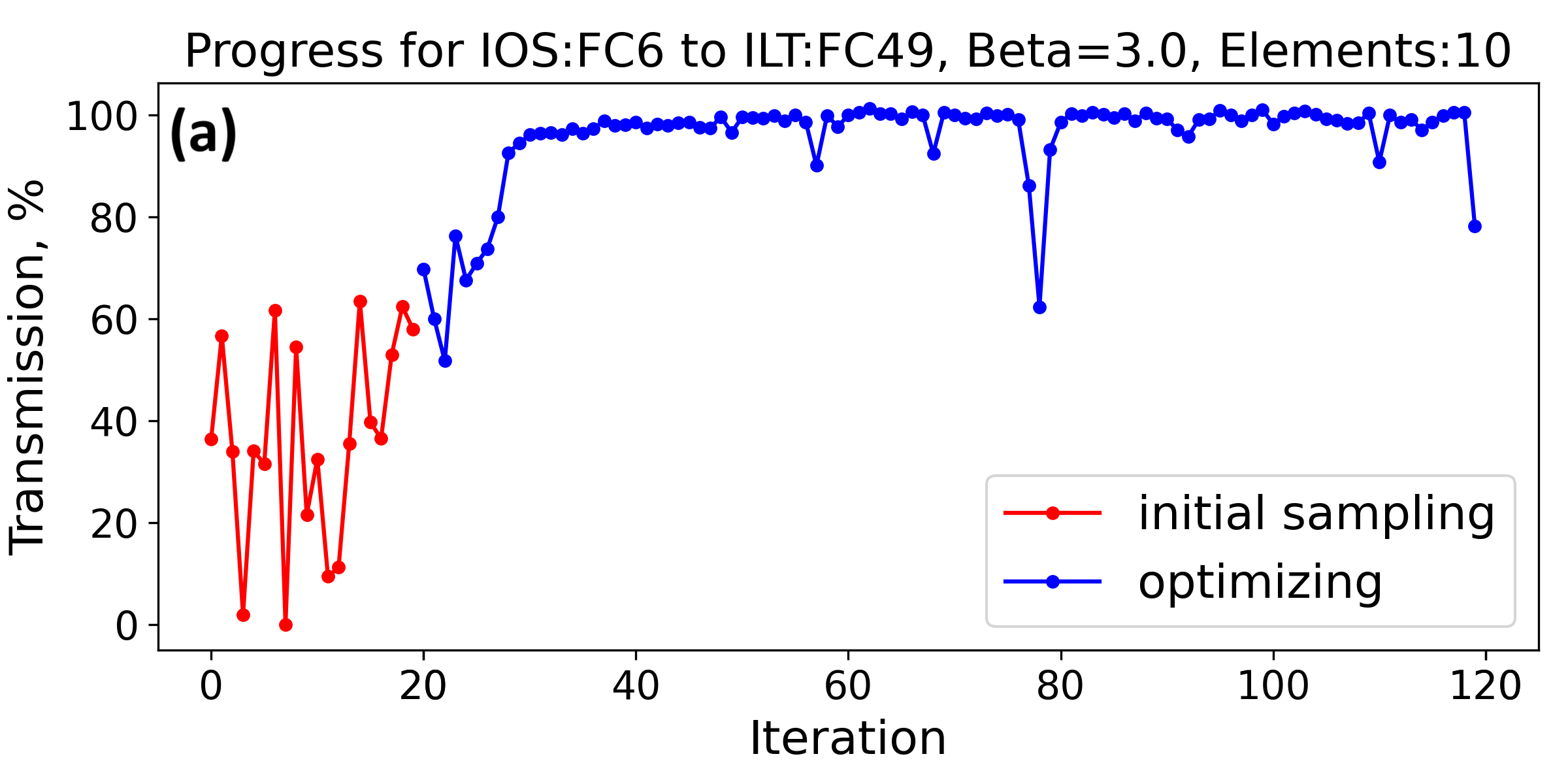}
    \end{subfigure}\hfill
    \begin{subfigure}{0.34\textwidth}
        \includegraphics[height = 3.2cm, keepaspectratio]{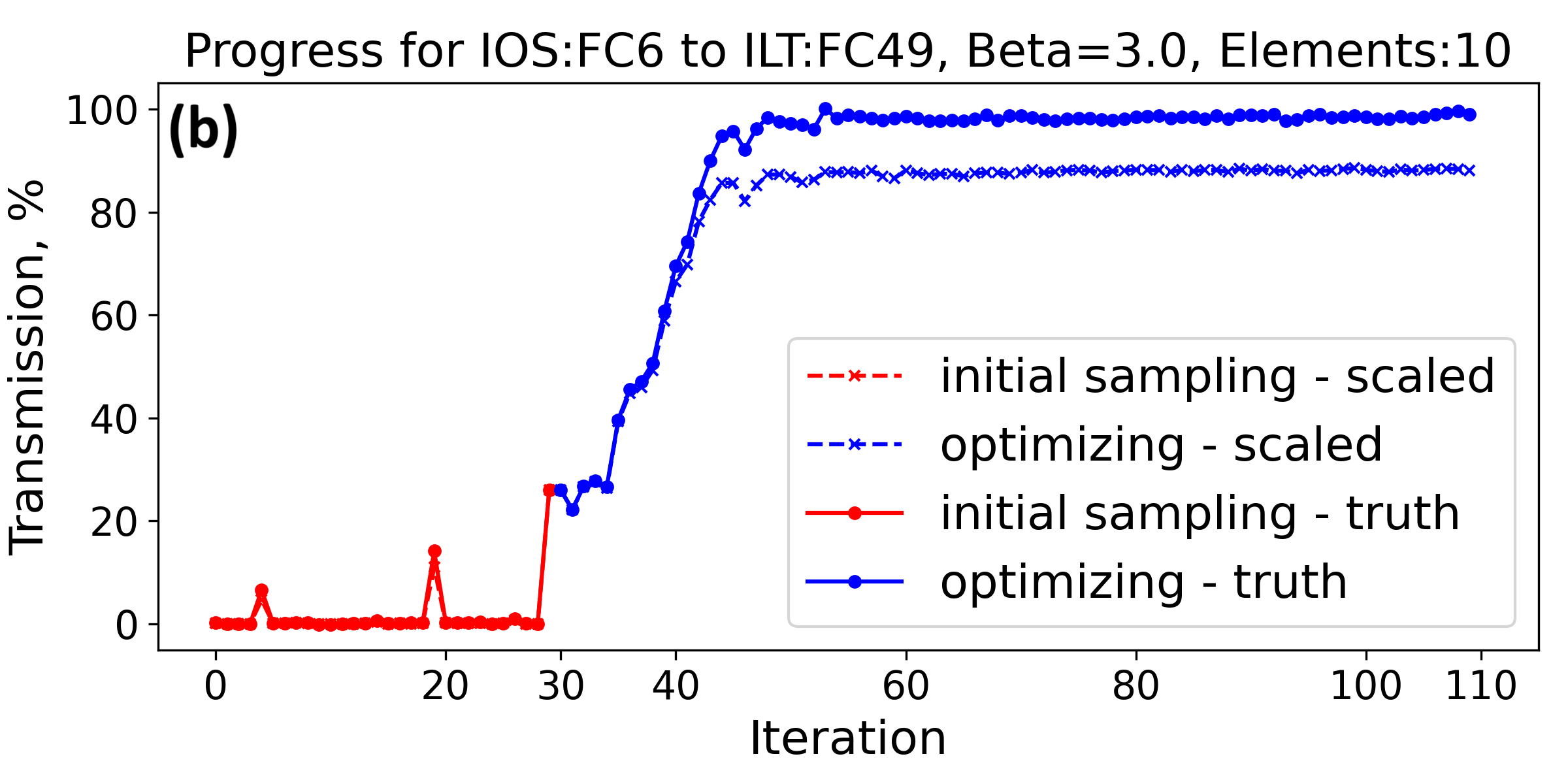}
    \end{subfigure}\hfill
    \begin{subfigure}{0.29\textwidth}
        \includegraphics[height = 3.5cm, keepaspectratio]{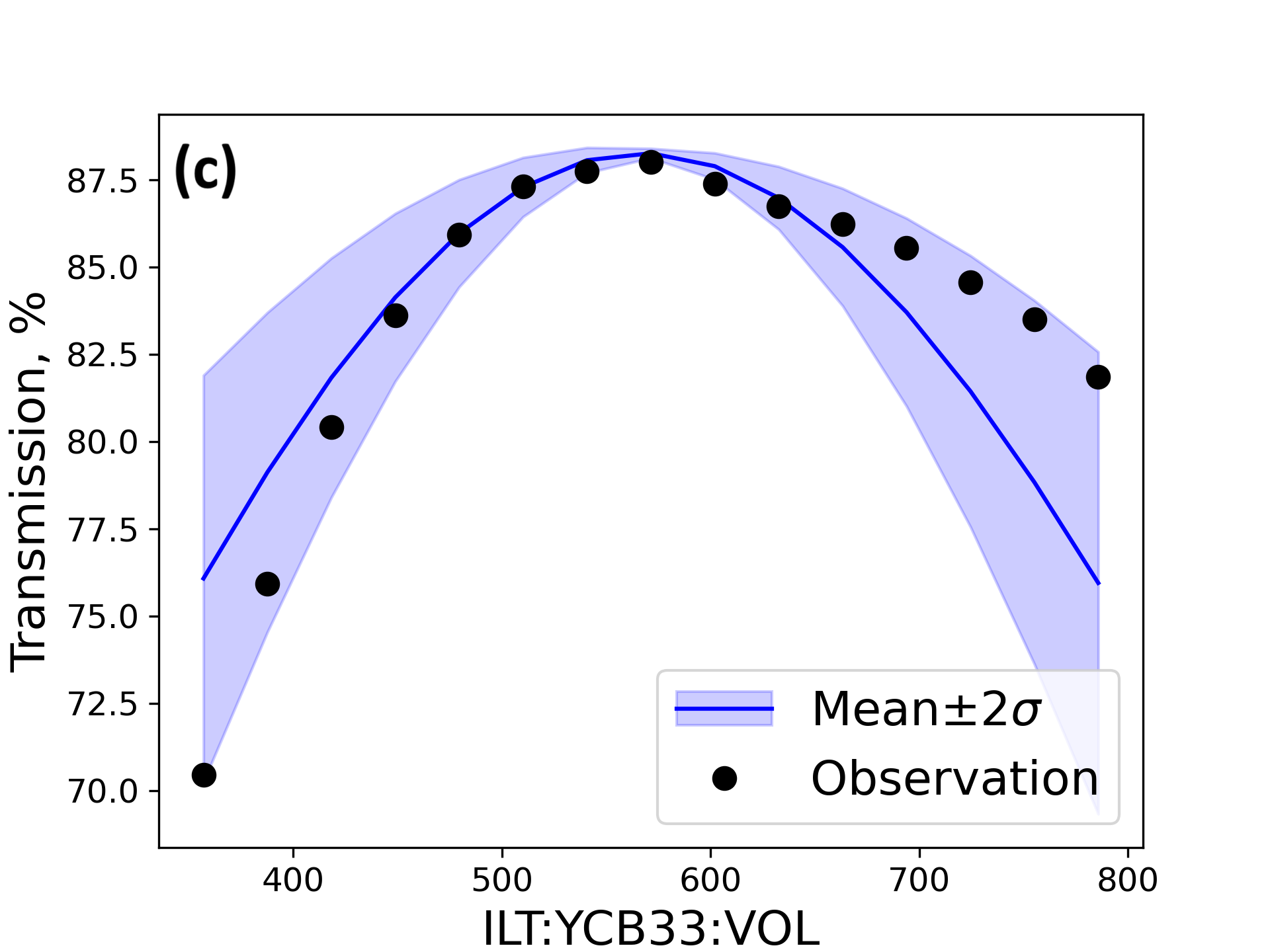}
    \end{subfigure}
\caption{Example of operator feedback plots created after optimization is complete. Plots (a) and (b) show progress divided into two regions for initial sampling and optimizing for two separate runs, and additionally plot (b) shows both the real and model-perceived values when using \texttt{scaleBOIS}. Plot (c) shows a 1D posterior scan as an example of what a user sees after optimization is complete, and is used to judge the proficiency of the model. Shown is the scan for only one dimension. The shaded region is the 95\% confidence region and the solid line is the mean, both of the posterior distribution. }
\label{fig:example}
\end{figure*}

Beam was transported from the IMS, through the low energy transport (LEBT) section, into the polarizer \cite{Kiefl2003bnmr},  and through the polarizer beamline (\autoref{fig:ISAC}). Additionally, OLIS beam was optimized by \texttt{BOIS} through the RFQ \cite{Marchetto2007postaccel} and accelerated into the MEBT section (\autoref{fig:ISAC}). Different beam compositions were used based on availability of ion beams during test times including \ce{^{7}Li+}, \ce{^{12}C+}, and \ce{^{22}Ne^{4+}}. The \texttt{scaleBOIS} and \texttt{boundBOIS} methods were both tested on-line and shown to be effective at optimizing transmission at the same level as the operators, while also reducing overall steering.

\subsection{Bayesian Optimization Beamline Tuning}
The first section shown starts from the mass separator magnet after the ISAC targets and goes to the polarizer in the experimental hall. The second section starts from the separator magnet after the OLIS and goes to the RFQ, first terminating at the linac’s entrance, then going through to the MEBT section.

During development and testing, before acquiring data on the \texttt{BOIS} method, an operator tuned the beamline steerers for maximum transmission per standard procedures. Once this was  completed, and the transmissions measured, the steerers were reset to neutral and the optimizer was executed. The manually established transmissions allow for a baseline comparison of the performance, compared to the typical manual tuning methodology.

At the end of the algorithm, two plots are produced: the model's progress at each iteration, and the scan results compared to the model's fit. User feedback plots are produced as results, and examples are shown in \autoref{fig:example}.

\subsubsection{\label{sec:IMS-Pol results}IMS - Polarizer}
The beamline from the IMS to the polarizer was broken into four subsections shown below (first being repeated). The optimized beam is \ce{^{7}Li+} at 25keV.

\begin{table}[h!]
\caption{Comparison of transmissions (tx) of online beam from operator and BO tunes.}
\begin{center}
\begin{tabularx}{0.49\textwidth}{
    >{\hsize = 2.1 \hsize \raggedright\arraybackslash}X 
    >{\hsize = 0.69 \hsize\centering\arraybackslash}X  
    >{\hsize = 0.69 \hsize\centering\arraybackslash}X  
    >{\hsize = 0.69\hsize\centering\arraybackslash}X  
    >{\hsize = 0.69 \hsize\centering\arraybackslash}X}
    \hline
    \hline
    Section (FC-FC) [letters reference \autoref{fig:ISAC}]& Section Length (m) & Used / Total Steerers & Operator tx (\%) & \texttt{BOIS} tx (\%) \\
    \hline
    \vspace{1ex}\\
    IMS:14 (a) - IMS:34 (b) & 9 & 13/13 &  80 & 73\\ 
    IMS:14(a) - IMS:34(b) & 9 & 4/13 &  80 & 89\\ 
    IMS:34(b) - ILE2:1(c)  & 15 & 17/21 &  $98^*$ & 95\\
    ILE2:1(c) - ILE2:11(d)  & 3 & 4/4 &  $98^*$ & 91 \\
    ILE2:11(d) - ILE2:19(e)  & 4 & 7/7 &  $86^*$ & 75 \\
    \vspace{1ex}\\
    \hline
    \hline
\end{tabularx}
\label{tab:IMS-POL_results}
\end{center}
$^*$ Note that operators tuned sections (b)-(e) with an extra 58\% attenuation at slits placed in section (a)-(b), which was not present in the BOIS tune, or the operator tune (a)-(b). However, no adverse effect is expected downstream of point (b), since in all cases, 2rms beamsize in the system is expected to be below aperture constraints. 
\end{table}

\begin{figure}[hb!]
    \centering
    \includegraphics[width = 0.49\textwidth]{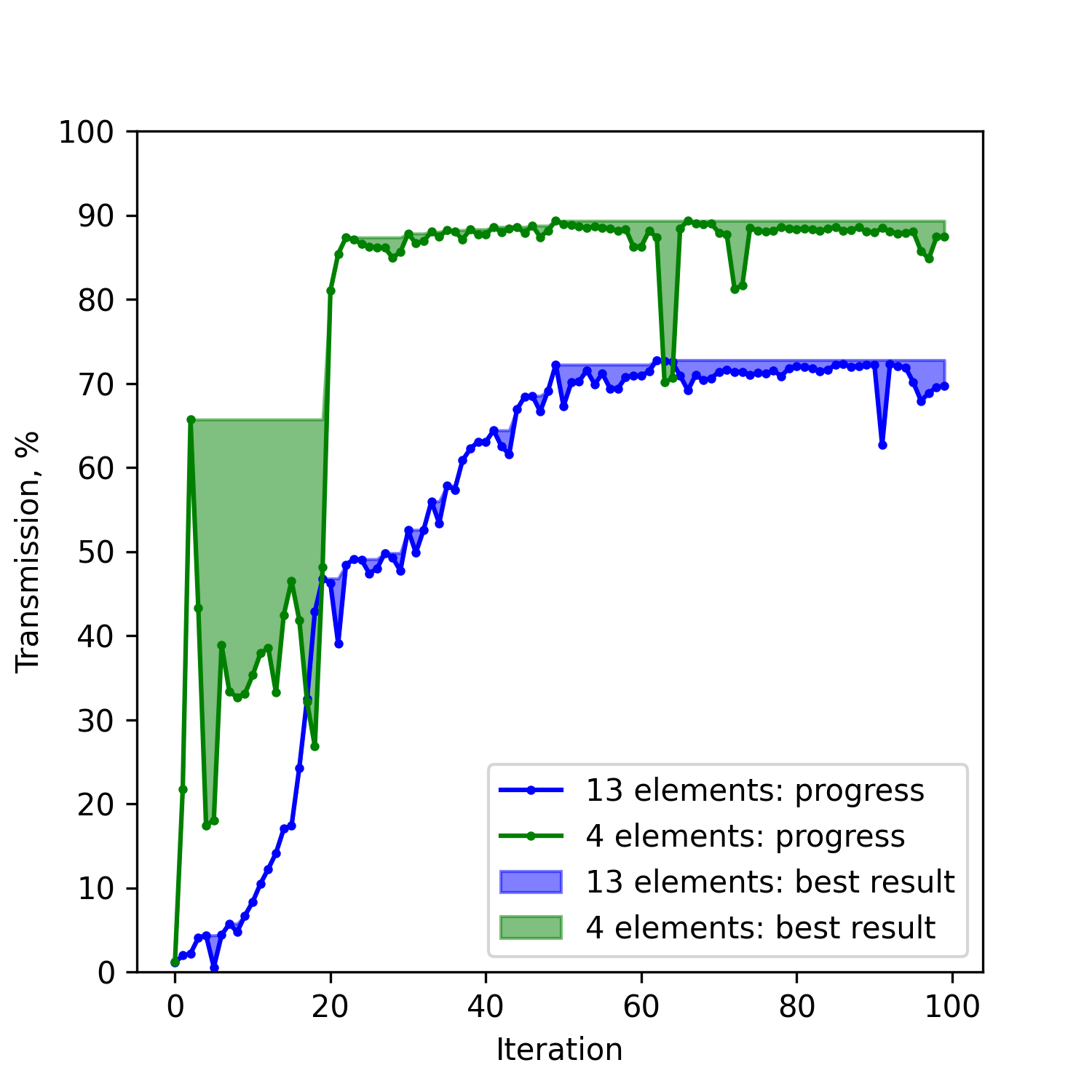}
    \caption{Comparison of transmission by two models, one with access to all elements in the section (\autoref{fig:ISAC} labels (f)-(g)) and one with access to carefully chosen elements. Line shows model progress through optimizing, and shaded region shows best achieved result.  This shows that selecting steerers improves both learning time and overall success.}
    \label{fig:convergance_comparison}
\end{figure}

The first section shown in \autoref{tab:IMS-POL_results}, from IMS:FC14 to IMS:FC34, was tuned by an operator to achieve 80\% transmission as a benchmark value before the BO algorithm achieved 73\%. Immediately upstream of this section is the mass separator magnet, which selects different masses produced in the ISAC targets with a mass resolution $M/(\delta M)$ of up to 2000. The section tuned here includes two set of slits to define the transverse beamsize by cutting, if required. Using this knowledge, an expert operator would  focus instead on the steering upstream of the slits, because they are the "pinch points" where most loss occurs. Taking this into account, the \texttt{BOIS} was used with only steerers upstream of the slits, a transmission of 89\% was achieved. These are the first two entries in \autoref{tab:IMS-POL_results}, and a comparison of transmission through iterations is shown in \autoref{fig:convergance_comparison}. 

Further down, in the final section shown in \autoref{tab:IMS-POL_results}, \texttt{BOIS} transports beam through the polarizer, which includes an aperture of 8 mm, that limits the maximum transmission around the range of 80\%.

\subsubsection{OLIS - RFQ}
\begin{figure}[hb]
    \centering
    \includegraphics[width = 0.49\textwidth]{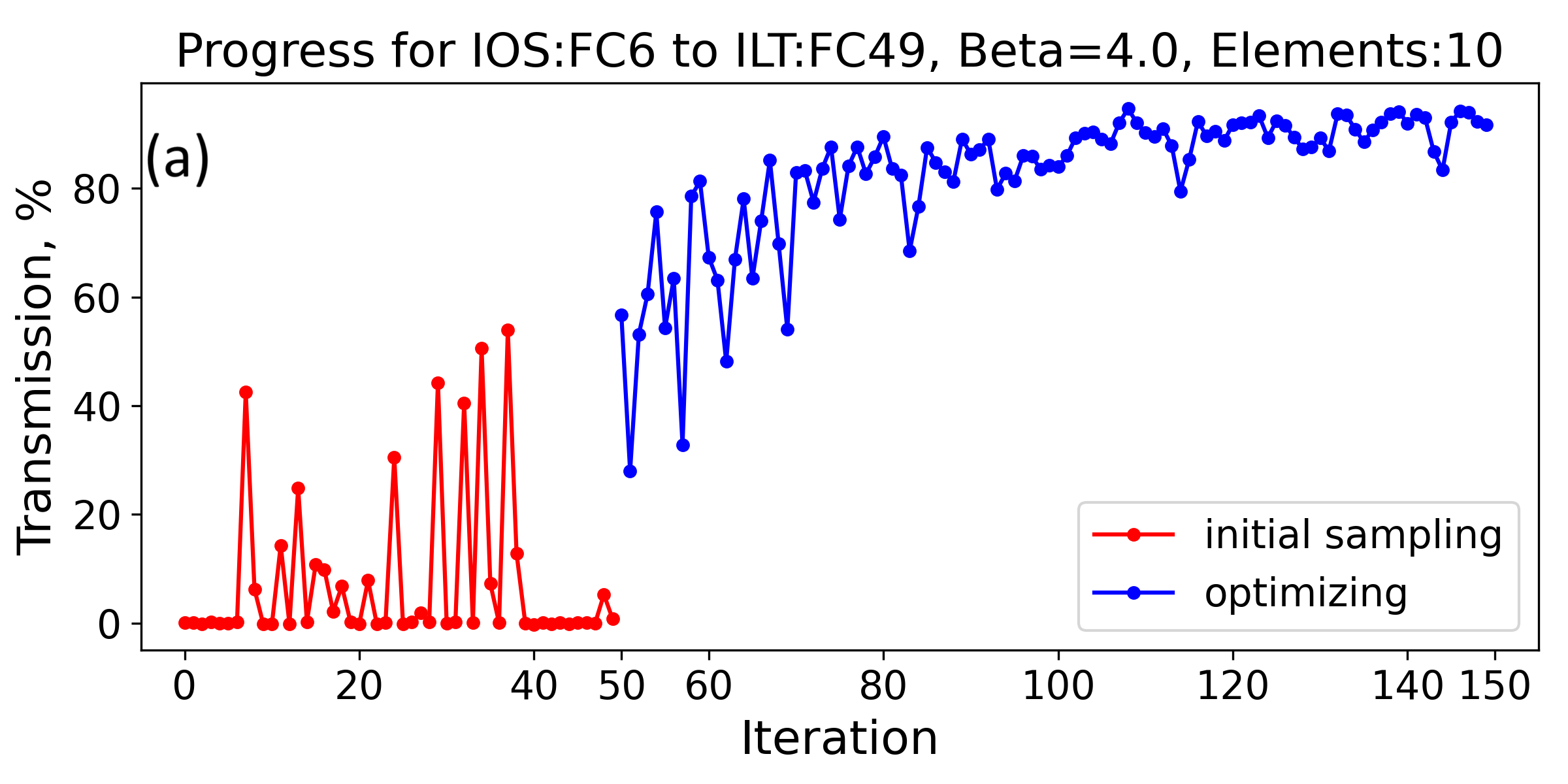}
    \includegraphics[width = 0.49\textwidth]{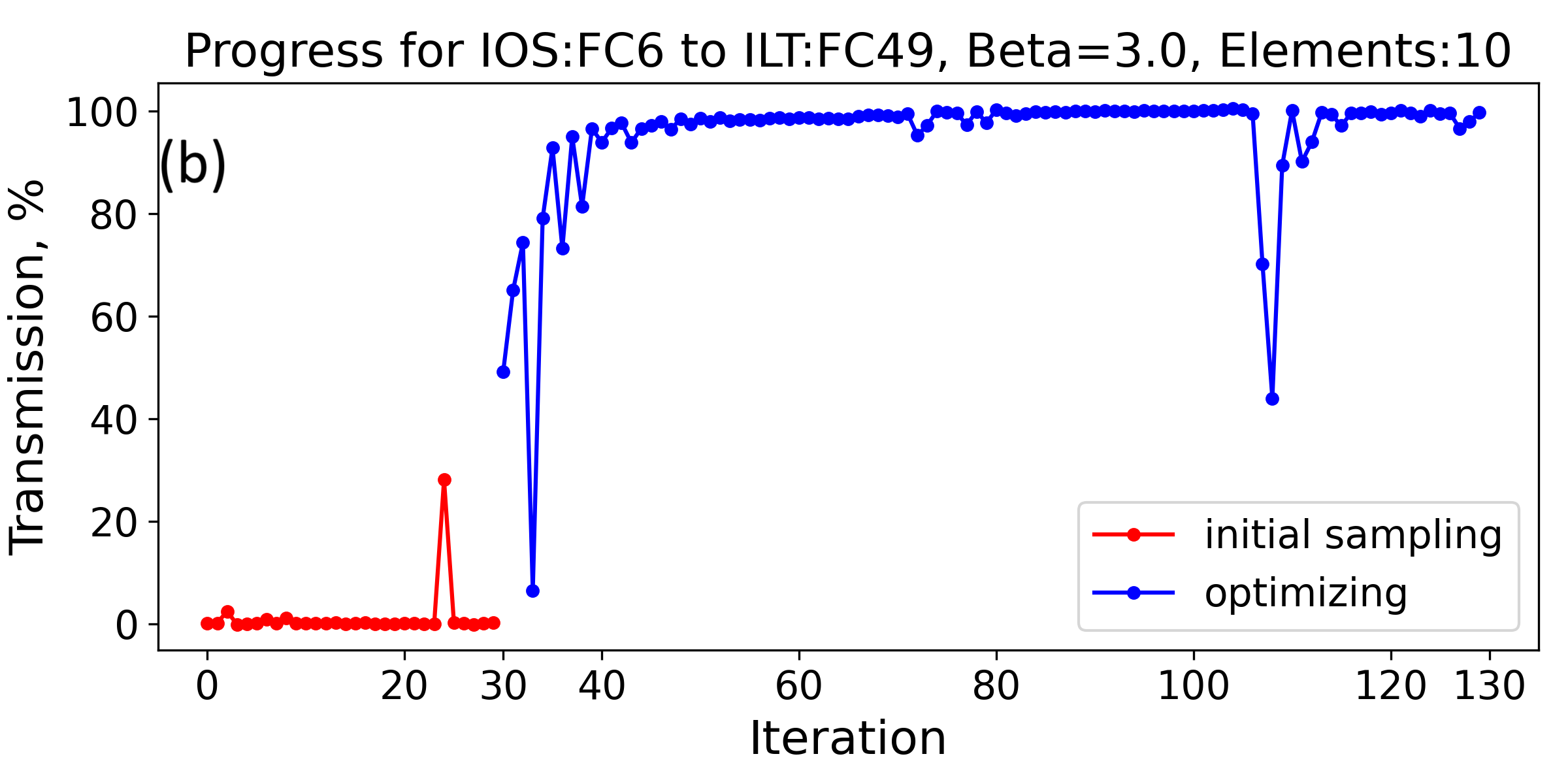}
    \caption{Comparing model behavior with different $\beta$ values using transmission of the beam transport from OLIS to the entrance of the RFQ. Two different beam species are shown: (a) \ce{^{22}Ne^{4+}} with an incoming current of 22.0~nA, and a final transmission of 95\%; (b) \ce{^{12}C+} with an incoming current of 7.92~nA, and a final transmission of 100\%.}
    \label{fig:OLIS_results} 
\end{figure}

OLIS houses three separate ion sources which can be selectively used to supply stable beams to the experiments at ISAC (\autoref{fig:ISAC}). In \autoref{tab:OLIS_results} results are shown for beams from two different sources, the microwave source\cite{jayamanna2008off} (MWS) for \ce{^{12}C+} and the multicharge ion source\cite{jayamanna2010multicharge} (MCIS) for \ce{^{22}Ne^{4+}}. All OLIS-RFQ data measured at 2.04~keV/u, where u denotes atomic mass units, which is the energy per nucleon required for the RFQ injection \cite{shelbaya2019fast}. \autoref{fig:OLIS_results} shows data for two different runs on this section, using \texttt{BOIS} on 10 steerers and showing the different behavior the model has with different values of $\beta$. Our acquisition function (UCB), which selects the next point to explore (\autoref{sec:BO}), balances exploring unknown regions with exploiting known maxima using the $\beta$ parameter. The initially erratic behavior plotted in \autoref{fig:OLIS_results} (a) is indicative of the algorithm's priority to choose regions with high uncertainty for measurement using $\beta=4$. Ultimately the search reaches a plateau, but far more slowly than in (b), which hones in to the success it has found using $\beta=3$ . 

\begin{table}[ht!]
\caption{Comparison of transmissions of OLIS beam from operator and BO tunes (section from IOS:FC6 to ILT:FC49, labels (f) to (g) in \autoref{fig:ISAC})}
\begin{center}
\begin{tabularx}{0.43\textwidth}{
    >{\hsize=2\hsize\linewidth=\hsize \raggedright\arraybackslash}X >{\hsize=0.5\hsize\linewidth=\hsize \centering\arraybackslash}X  >{\hsize=0.5\hsize\linewidth=\hsize \centering\arraybackslash}X}
    \hline
    \hline
    Isotope & Operator tx (\%) & \texttt{BOIS} tx (\%) \\
    \hline
    \vspace{1ex}\\
    \ce{^{22}Ne^{4+}} & 85 & 95\\ 
    \ce{^{12}C+} & 100 & 100 \\
    \vspace{1ex}\\
    \hline
    \hline
\end{tabularx}
\label{tab:OLIS_results}
\end{center}
\end{table}

Using the same procedure, except including 3 additional steerers before the RFQ, the transmission was measured at the high energy end of the RFQ linac. Following standard procedure, the longitudinal tune of the RFQ, including multiharmonic \cite{poirier1999rf} pre-buncher time-focus into the linac, and vane voltage settings, was done manually by operators.

\begin{figure}[!t]
    \centering
    \includegraphics[width = 0.49\textwidth]{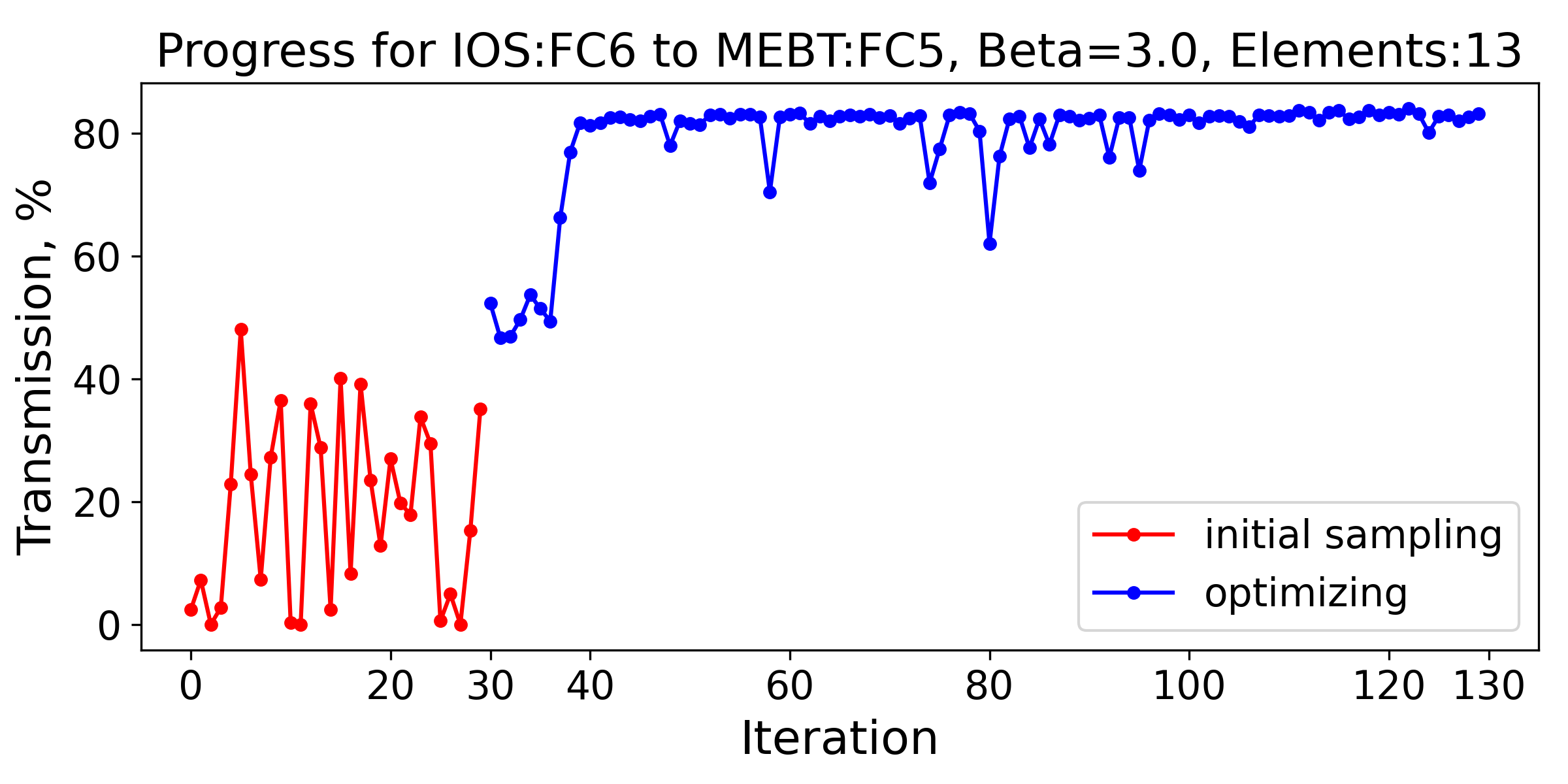}
    \caption{Transmission of the beam transport from the stable ion source, OLIS, to five quads after the exit of the RFQ, \autoref{fig:ISAC} label (f) to (h).  Beam is \ce{^{12}C+} with an incoming current of 7.28nA, and a final transmission of 84\%.}
    \label{fig:RFQ_12C_results}
\end{figure}
The RFQ transmission obtained by \texttt{BOIS} is 84\%, as seen in \autoref{fig:RFQ_12C_results}, noting that the separated pre-buncher induces a longitudinal intensity modulation akin to a time-focus at the linac's entrance\cite{shelbaya2019fast}, limiting total transmission to the mid 80\% range \cite{Laxdal1998RFQtest}.

\subsection{Variations for Reduced Steering}
Efforts were also made toward getting solutions that had an additional focus of keeping steering minimal. This was done two ways: using the \texttt{scaleBOIS} scalarization multi-objective method, \autoref{sec:scaling}; and using the \texttt{boundBOIS} $\varepsilon$-constraint multi-objective method, \autoref{sec:bounds_method}. 

\begin{figure}[h!]
    \centering
    \includegraphics[width = 0.50\textwidth ]{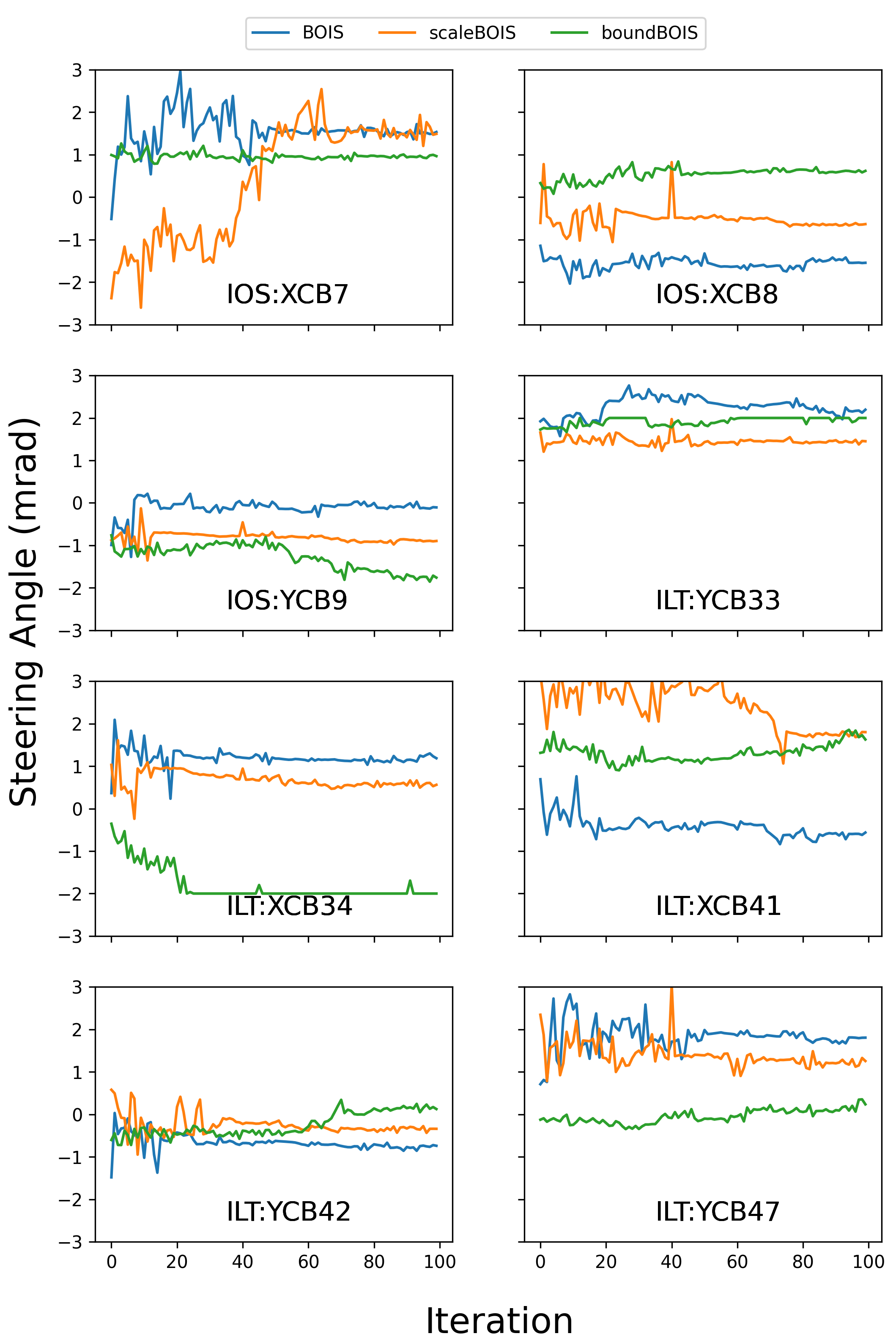}
    \caption{Steering angles explored for 10 steering elements from IOS:FC6 to ILT:FC49 (\autoref{fig:ISAC} label (f)-(g)), comparing scaled transmission input and enforcing bounds on steering. Plot begins after 50 random samples. All tests used $\beta=4$. The beam is \ce{^{22}Ne^{4+}}.}
    \label{fig:compare_all_models}
    
\end{figure}
\begin{table}
\caption{Summarized results from the run displayed in \autoref{fig:compare_all_models}. Values calculated as the absolute value mean of all steering applied. Results are only for one run of data.}
\begin{center}
\begin{tabularx}{0.3\textwidth}{
    >{\hsize=0.6\hsize\linewidth=\hsize \raggedright\arraybackslash}X  >{\hsize=1.4\hsize\linewidth=\hsize \centering\arraybackslash}X}
    \hline
    \hline
    \texttt{BOIS} type & mean abs final steering angle (mrad) \\
    \hline
    \vspace{1ex}\\
    \texttt{BOIS} & 1.05\\ 
    \texttt{scaleBOIS}   & 1.003\\
    \texttt{boundBOIS} & 0.78 \\
    \vspace{1ex}\\
    \hline
    \hline
\end{tabularx}
\label{tab:types&angles}
\end{center}
\end{table}

We compared the different variations using a \ce{^{22}Ne^{4+}} beam from MCIS, as shown in \autoref{fig:compare_all_models} and summarized the results in \autoref{tab:types&angles}. This figure shows the values that the models explore while optimizing.  The unbounded runs often explore outside of the reduced steering area ($\pm 2 \text{~mrad}$), but generally fall within it.
From the runs shown in \autoref{fig:compare_all_models}, the transmissions were 95\% for \texttt{BOIS}, 100\% for \texttt{scaleBOIS}, and 79\% for \texttt{boundBOIS}. This suggests that here the solution for this section lies outside of the $\pm2$~mrad constraint set by \texttt{boundBOIS}. This encourages more investigation into bounding constraints on different beamline sections and different source configurations. The results in \autoref{tab:types&angles} show that while \texttt{boundBOIS} was successful in reducing the overall steering, the strictness parameter of \texttt{scaleBOIS} was not high enough for a large reduction from \texttt{BOIS}. 

\subsection{Choosing a beamline subsection}
When using this tool on a new beamline, the elements to use for the optimization must be chosen first. Although \autoref{tab:IMS-POL_results} shows that selecting a few steerers can improve performance in cases with specific knowledge, this is often not possible and the general practise with \texttt{BOIS} is to utilize all available steering elements in a beamline. For example, the longest section of beamline we successfully optimized used 17 steerers, shown in \autoref{tab:IMS-POL_results}. A rough upper limit on the number of steerers in one section is 20, which is the current dimension of applicability for BO \cite{Frazier2018tutorial}.

\section{\label{sec:conclusion}Conclusion}
We have presented an application of Bayesian optimization to beam centroid steering. This method, \texttt{BOIS}, finds an optimal solution in relatively few function evaluations using noisy data. Further extensions of \texttt{BOIS}, \texttt{scaleBOIS} and \texttt{boundBOIS}, have been presented to reduce steering usage while still tuning for transmission using methods which reduce this multi-objective problem to a single objective. The tool outlined in this work was able to reliably transport beam with high transmission, matching operator performance. 
The results from tests with \texttt{boundBOIS} were the most promising, but further tests across more beamlines are required to explore constraints. Similarly, \texttt{scaleBOIS} was nominally effective, but the response to the strictness parameter needs further investigation.
The code, run on CPU and not optimized for efficiency, took comparable time to manual tuning. 

Looking ahead, we intend to both expand the usage of this program, and update its capabilities as part of TRIUMF's HLA. Currently, we have only extensively tested in the low-energy section of ISAC, but intentions are to expand this method from low-energy (electrostatic) to include high-energy (magnetic) steering. Investigations are being carried out in the higher energy areas downstream of the accelerator, which use magnetic lenses and steerers.

\section*{Acknowledgements}
We thank T. Planche and O. Hassan for useful questions and discussions. P. Jung and S. Kiy are thanked for assistance in the design of the control system communication protocols and general software expertise. Thanks to all ISAC RIB operators for shift assistance. This work is supported by the Natural Sciences and Engineering Research Council of Canada (NSERC), under grant no. SAPPJ-2023-00038 and RGPIN-2018-04030. We acknowledge that TRIUMF is located on the traditional, ancestral, and unceded territory of Musqueam people, who for millennia have passed on their culture, history, and traditions from one generation to the next on this site. 

\section*{References}

\providecommand{\noopsort}[1]{}\providecommand{\singleletter}[1]{#1}%

\end{document}